\documentclass[a4paper,11pt]{article}

\newcommand\mainmatter{%
\clearpage
\pagenumbering{arabic}
}

\usepackage{authblk}

\usepackage[T1]{fontenc} 

\usepackage[utf8]{inputenc}
\usepackage{amsmath,amsfonts,amsthm}
\usepackage{graphicx}
\usepackage{amsmath}
\usepackage{slashed}
\usepackage{comment}
\usepackage{color}
\usepackage{mathtools}
\usepackage{hyperref}
\numberwithin{equation}{section}
\usepackage[normalem]{ulem}
\usepackage[super]{nth}
\usepackage[top=2cm,bottom=2cm,outer=2.5cm,inner=2.5cm,marginparwidth=2.5cm]{geometry}

\newcommand\frontmatter{%
\clearpage
\pagenumbering{roman}
}

\usepackage{cite}

\usepackage{braket}
\usepackage{amssymb}

\usepackage{float}
\usepackage{mathrsfs}

\newcommand{\beq}{\begin{equation}}
\newcommand{\eeq}{\end{equation}}
\newcommand{\bal}{\begin{equation}\begin{aligned}{}}
\newcommand{\eal}{\end{aligned} \end{equation}}

\def\bR {\mathbb{R}}
\def\bZ {\mathbb{Z}}
\def\cN {\mathcal{N}}

\title{Vortex loop operators and quantum M2-branes}
\author{Nadav Drukker\footnote{\href{mailto:nadav.drukker@gmail.com}{nadav.drukker@gmail.com}} }
\author{Omar Shahpo\footnote{\href{mailto:omar.shahpo@kcl.ac.uk}{omar.shahpo@kcl.ac.uk}}}
\affil{\it Department of Mathematics, King's College London,\protect\\London, 
WC2R 2LS, United Kingdom}
\date{}

\begin{document} 

\frontmatter
\maketitle

\begin{abstract}

We study M2-branes in $AdS_4\times S^7/\bZ_k$ dual to 
1/2 and 1/3 BPS vortex loop operators in ABJM theory 
and compute their one-loop correction beyond the classical M2-brane action. 
The correction depends only on the parity of $k$ and is independent 
of all continuous parameters in the definition of the vortex loops. 
The result for odd $k$ agrees with the answers for the 1/2 BPS Wilson 
loop in the $k=1$ theory and for even $k$ with the one in the 
$k = 2$ theory. Combining with the classical part, we find that 
the natural expansion parameter seems to be $1/\sqrt{kN}$ rather 
than $1/\sqrt{N}$. 
This provides a further setting where semiclassical 
quantisation can be applied to M2-branes and produces new 
results inaccessible by other methods.

\end{abstract}
\thispagestyle{empty}

\mainmatter
\tableofcontents

\section{Introduction and conclusions}
\label{sec:intro}

The holographic duality between M2-branes and observables in conformal 
field theories 
\cite{Maldacena:1997re, Aharony:2008ug} enables a 
deeper understanding of both. Field theory techniques such as 
localisation \cite{Pestun:2007rz, Kapustin:2009kz} provide exact 
predictions for M2-brane partition functions. Those can be matched 
with classical M2-brane computations~\cite{Drukker:2008zx, 
Chen:2008bp, Rey:2008bh,Cagnazzo:2009zh, Drukker:2010nc, 
Drukker:2011zy, Hatsuda:2013gj, Gautason:2023igo} 
and more recently semiclassical calculations 
\cite{Giombi:2023vzu, Beccaria:2023hhi, Beccaria:2023ujc, Beccaria:2023sph}.

Conversely, in cases when the field theory techniques are not as 
well developed, the M2-brane description may fill the gap by 
providing strong coupling results. 
For surface operators in the 6d (2,0) theory \cite{Ganor:1996nf}, 
M2-branes are crucial in computing their expectation 
values to leading~\cite{Maldacena:1998im, Berenstein:1998ij, 
Graham:1999pm, Drukker:2020dcz} 
and subleading~\cite{Drukker:2020swu, Drukker:2023jxp} 
orders at large $N$. 
In ABJM theory, M2-branes provide the holographic description of 
a class of 1d observables known as vortex loops \cite{Drukker:2008jm}. 
In this paper, we extend the computation of their expectation values 
to one-loop using semiclassical M2-brane tools.

Vortex loops are operators supported on a closed contour 
on which some of the matter and gauge fields are singular. 
They are disorder operators, similar to 't~Hooft or surface operators in 4d 
gauge theories. 
In pure Chern-Simons theory, they are equivalent to Wilson 
loops~\cite{Moore:1989yh}, while they have richer features in 
Chern-Simons-matter and other interacting 3d theories. 
They were described in ABJM theory in~\cite{Drukker:2008jm} and 
the mapping between them and Wilson loops under 3d $\cN=4$ mirror symmetry 
was explained in \cite{Assel:2015oxa}. Particular examples of abelian 
vortices in $\cN=2$ were computed using supersymmetric localisation 
in~\cite{Kapustin:2012iw, Drukker:2012sr}.

We study the holographic description of 1/2 and 1/3 BPS vortex loops in ABJM 
theory. The simplest class are characterised on the field theory side by the 
gauge symmetry breaking to $U(N-1)^2 \times U(1)$ \cite{Drukker:2008jm}. 
In coordinates $(t, z,\bar{z})$ on $\mathbb{R}^3$ with $z$ a complex coordinate 
on the spatial slice, placing the vortex line at $z=0$, the four ABJM bifundamental 
matter fields $C^I$ take the singular classical configuration 
\begin{equation}\label{Eq:Singularity}
C^{1} = \frac{1}{\sqrt{z}}
\begin{pmatrix}
0_{N-1} & 0\\
0 & \beta_1
\end{pmatrix}\,,
\qquad
C^{2} = \frac{1}{\sqrt{\bar{z}}}
\begin{pmatrix}
0_{N-1} & 0\\
0 & \beta_2
\end{pmatrix}
\,,
\qquad
C^{3} = C^{4} = 0\,,
\end{equation}
where $\beta_{1,2}$ are complex parameters, while the two gauge fields of ABJM $A$ and $\hat{A}$ are given by
\begin{equation}
A_z = \hat{A}_z = -\frac{i}{4kz}\begin{pmatrix}
0_{N-1} & 0\\
0 & \alpha
\end{pmatrix}\,,
\qquad
A_t = \hat{A}_t = - 2\pi(C^1 C_1^\dagger-C^2 C_2^\dagger)\,,
\end{equation}
where $\alpha$ is an angular variable. This configuration is 
1/3 BPS, with supersymmetry enhanced to 1/2 when $\beta_2=0$.

At the quantum level, a vortex loop is defined by the path integral 
with this singular behaviour near $z=0$. The definition above is for the 
infinite straight line, but to get a finite expectation value we should 
really look at the circle, which is just a conformal transformation of 
the above configuration, as is familiar in the case of Wilson loops 
\cite{Drukker:2000rr}.

The vortex loop \eqref{Eq:Singularity} is described in 
$AdS_4\times S^7/\mathbb{Z}_{k}$ by a single M2-brane with 
$AdS_2\times S^1$ geometry \cite{Drukker:2008jm}. For the 
background we take the metric
\begin{equation}\label{background metric}
ds^2 = \frac{R^2}{4} \left(du^2+\cosh^2u\,h_{AdS_2}
+\sinh^2u \, d\phi^2\right)
+ R^2 ds^2_{S^7/\mathbb{Z}_k}\,,
\end{equation}
and $S^7/\mathbb{Z}_k$ can be written as a $U(1)$ fibre over $\mathbb{CP}^3$, 
see \eqref{S7 metric}. This choice of metric is an $AdS_2\times S^1$ foliation 
of $AdS_4$ and its boundary at $u\to\infty$ has the geometry $AdS_2\times S^1$. 
This metric is conformal to $\bR^3$ such that under the conformal 
transformation, the boundary of $AdS_2$ is mapped to 
a line or a circle. Therefore, this metric is ideal for studying 
conformal circle and line operators. 
More specifically, the boundary of $AdS_2$ 
for all values of $u$ are mapped to the same line/circle.

The M2-branes are extended along $AdS_2$ at fixed $u=u_0$. The third direction 
wraps the $\phi$ circle while also following the fibre direction 
(and in the case of the $1/3$ BPS loop, also a circle on the $\mathbb{CP}^3$ 
base). All cycles are mutually periodic when the fibre direction is wrapped 
$k$ times and the others are wrapped twice. For even $k$ it suffices then 
to have $k/2$ periods, which was shown in \cite{Drukker:2008jm} to match 
the properties of the field theory vortex at even $k$.

This holographic description was used in \cite{Drukker:2008jm} to compute the 
vortex loop expectation value and some correlation functions at strong coupling. 
As reviewed in Section~\ref{sec:classical}, 
the answer is given by the classical M2-brane action
\begin{equation}\label{Eq:ExpectationValue}
\log \langle V \rangle = 
-S^{(0)}
=
\begin{cases} 
2\pi \sqrt{kN/2} + O(N^0)& k \text{ odd.}\\
\pi \sqrt{kN/2} + O(N^0) & k \text{ even.}
\end{cases}
\end{equation}
The result is independent of $u_0$ (or $\beta_1$ in field theory language, 
c.f. \eqref{Eq:ParameterMap}) 
and is also the same for any of the $1/3$ BPS loops (so arbitrary $\beta_2$). 
In the limit of $u_0=0$ the brane configuration agrees with that describing 
$k/2$ or $k$ coincident Wilson loops. 
Unsurprisingly, the classical answer is $k/2$ or $k$ times the answer 
for the 1/2 BPS Wilson loop \cite{Drukker:2009hy}.

More precisely, the $k$ coincident M2-branes should correspond to 
a Wilson loops in some (possibly reducible) $k$-dimensional representation 
of $U(N)$ (or really $U(N|N)$ for the 1/2 BPS loop \cite{Drukker:2009hy}). 
Recall that monopole operators in Chern-Simons theories transform in 
the $k$-dimensional symmetric representation~\cite{Borokhov:2002ib, 
Borokhov:2002cg, Aharony:2008ug}. This has the effect that charge 
can change by $k$ units and is 
sometimes referred to as the bosonic exclusion 
principle~\cite{Minwalla:2020ysu}. This may be behind the fact that the 
$k$-dimensional Wilson loop may be continuously deformed into a 
vortex loop with arbitrary $u_0$.

To compute the one-loop correction to the classical action 
\eqref{Eq:ExpectationValue}, we study the M2-brane's quadratic fluctuation action. 
This follows on the study of fluctuations of classical string solutions 
in $AdS_5\times S^5$ \cite{Drukker:2000ep}, and 
of M2-branes in \cite{Forste:1999yj,Sakaguchi:2010dg, Drukker:2020swu}. 
This is the main computation of this paper and is in Sections~\ref{Sec:halfBPS} 
and~\ref{Sec:ThirdBPS} below.

We find that the quadratic action is identical in both the 1/2 and 1/3 BPS 
cases and depends on the parameter $u_0$ only via overall prefactors. 
One can absorb these factors by rescaling the fluctuating fields and 
in any case, it does not affect the computation of the one-loop 
determinant. The action \eqref{quad-action}, \eqref{fermion action} 
contains explicit dependence on $k$, but this can be removed by rescaling 
the coordinate $\zeta$ and bringing its range back to $[0,2\pi]$. 
So the answer is really also insensitive to the value of $k$, only to whether 
it is even or odd.

We therefore find that as with the classical calculation, 
the quadratic action agrees with that for the 1/2 BPS Wilson loop in the 
$k=1$ and $k=2$ theories, as derived in \cite{Sakaguchi:2010dg}. The 
determinants of these actions were evaluated in \cite{Giombi:2023vzu} 
by Kaluza-Klein reduction on the $S^1$ circle and using known results 
for determinants on $AdS_2$~\cite{Drukker:2000ep, Kim:2012tu, 
Buchbinder:2014nia, Giombi:2020mhz}. 
The results for the single Wilson loop 
depends on the value of $k$, but we only need to borrow the results 
for $k=1,2$ which are respectively $-\log4$ and 0. So we find 
that \eqref{Eq:ExpectationValue} gets corrected at one-loop to
\begin{equation}\label{Eq:ExpectationValueOneLoop}
\log \langle V \rangle = 
\begin{cases} 
2\pi \sqrt{ k N/2} - \log 4 + O(N^{-1/2})\,, & k \text{ odd,}\\
\pi \sqrt{k N/2} + 0 + O(N^{-1/2})\,, & k \text{ even.}
\end{cases}
\end{equation}

In the absence of an exact computation of the expectation value of the 
vortex loop we may wish to learn more general lessons from the classical 
and semiclassical answer.

First, we see that there is no dependence on $u_0$ and it is natural to 
conjecture that this persists to all orders in the $1/\sqrt{N}$ expansion. 
$u_0$ does not appear at all in the classical action and can be absorbed into 
the definition of the quadratic fluctuation fields. If a similar mechanism 
extends to the interacting theory, there will be no $u_0$ dependence.

Second, the dependence on $k$, apart for the separation into odd 
and even cases appears together with $N$, so the natural expansion 
parameter seems to be $1/\sqrt{kN}$, rather than $1/\sqrt{N}$. This 
combination is related to the radius of space (see \eqref{matching}) 
and is insensitive to the 
order of the orbifold $k$. This is not surprising, as the solution 
wraps the orbifold $k$ or $k/2$ times, so is natural to study it in 
the covering space. 
Indeed, for this perturbative calculation, it is not clear how 
the order of the orbifold could show up, as for $u_0\neq0$ the 
M2-brane never crosses itself and the local geometry is always like 
in the $k=1,2$ case.

If we focus on the theories with $k=1,2$, those have enhanced 
$\cN=8$ supersymmetry and the vortex loop 
preserves 16 supercharges. In fact, the 1/3 BPS loop is now 
enhanced to 1/2 BPS, as $SO(8)$ symmetry allows to rotate 
$C_2^\dagger$ into $C^1$. Then, if indeed the $k$ dependence 
factors with $N$, this result extends to all $k$.

It is natural to conjecture that vortex loops are described exactly by 
the 1/2 BPS Wilson loop in the $k=1,2$ theory, or that they differ by 
non-perturbative corrections. This requires either exact field theory 
calculations or the study of the interaction of the M2-branes 
describing the vortex with M2-brane instantons.

Unfortunately, $k=1,2$ are the only cases when the exact expectation 
value of the 1/2 BPS circular Wilson loop is not known. 
The expectation value of the Wilson loop was evaluated 
from the localisation matrix model at finite $k$ and large $N$ 
in~\cite{Herzog:2010hf}. This was extended to all orders in $N$ 
in~\cite{Klemm:2012ii} (see also~\cite{Marino:2009jd, 
Marino:2011eh, Fuji:2011km}). Unfortunately, the resulting expression 
diverges for $k=1,2$. Therefore, there is no way to try to 
borrow the Wilson loop result and try to make a conjecture for 
the vortex loop. A more careful localisation calculation 
is required for $k=1,2$ and this would hopefully supply the 
exact answer for both the Wilson and vortex loop.

\section{The 1/2 BPS configuration}
\label{Sec:halfBPS}

The vortex operators \eqref{Eq:Singularity} are described by single M2-branes 
in $AdS_4\times S^7/\mathbb{Z}_k$, with the metric \eqref{background metric} 
in the introduction.%
\footnote{For the indices, we use $\mu,\nu$ to indicate $AdS_4$ indices, 
$a,b$ for $AdS_2 \subset AdS_4$, $m,n$ for  ${\mathbb{CP}}^3$ 
indices, and 7 for the $U(1)$ fibre. $A,B$ are used for the full 11d 
space and $i,j$ for the 3d world-volume. We employ hats to indicate 
flat tangent space indices.}
We write the metric on $S^7/\mathbb{Z}_k$ as a $U(1)$ bundle over ${\mathbb{CP}}^3$
\bal\label{S7 metric}
ds^2_{S^7/\mathbb{Z}_k} &= ds^2_{{\mathbb{CP}}^3} + \frac{1}{k^2}(k A + d\zeta)^2\,, \\
ds^2_{{\mathbb{CP}}^3} &= \frac{(1 + w^m\bar{w}^{\bar m}) dw^{n} d\bar{w}^{\bar{n}} - \bar{w}^{\bar{n}} w^m dw^n d\bar{w}^{\bar{m}}}{(1 + w^k\bar \bar{w}^{\bar k})^2}\,.
\eal
where $w^{n}$ are complex coordinates, with the bar denoting complex conjugation,
and $\zeta$ has $2 \pi$ period. $A$ is a real one form
\begin{equation}\label{Eq:OneForm}
A = -\frac{i}{2} 
\frac{\bar{w}^{\bar{n}} dw^{n} -w^{n} d\bar{w}^{\bar{n}}}
{1 + w^{m} \bar{w}^{\bar{m}}}\,.
\end{equation}
The K\"ahler form on $\mathbb{CP}^3$ is then $K=dA/2$.

The background also has a four-form $F_4$ proportional to the volume 
form on $AdS_4$, with a potential $C_3$, such that
\begin{equation}\label{Eq:C3}
C_3 = \frac{R^3}{8}(\cosh^3 u -1) \Omega_{AdS_2}\wedge d\phi\,,
\qquad
F_4 = dC_3 = \frac{3R^3}{8} \Omega_{AdS_4}\,.
\end{equation}
The choice of gauge for $C_3$  is compatible with the $AdS_2$ symmetry of the 
problem. This is actually crucial in the calculation of both the classical 
action and the fluctuations due to the fact that the brane is non-compact.

The relation between the gravity quantities and those of the field theory are
\beq
\label{matching}
\frac{R^3}{l_{\text{pl}}^3} = 4 \pi\sqrt{2kN}\,,
\eeq
where $l_{\text{pl}}$ is Planck length.

\subsection{Classical M2-brane solution}
\label{sec:classical}

The classical M2-brane action \cite{Bergshoeff:1987cm}
has a Nambu-Goto part and a Wess-Zumino part coupling 
the embedding of the M2-brane to the pullback (indicated by a star) 
of the 3-form $C_3$
\begin{equation}
\label{classical action}
S = T_{\text{M2}} \int\left( \Omega_{\text{M2}} +{}^*C_3\right),
\end{equation}
where $\Omega_{\text{M2}}$ is the volume form on the M2-brane. 
$T_{\text{M2}}= {1}/{4\pi^2 l_{\text{pl}}^3}$ is the M2-brane tension. 

The M2-brane dual to the 1/2 BPS vortex \eqref{Eq:Singularity} with 
$\beta_2 = 0$ is extended in $AdS_2$ at fixed $u_0$. The remaining 
direction is a circle embedded in both $AdS_4$ and $S^7/\bZ_k$. 
Parameterising it in terms of $\zeta$ from \eqref{S7 metric}, 
the solution is \cite{Drukker:2008jm}
\begin{equation}\label{classical}
u = u_0\,, \qquad 
\phi = \frac{2}{k}\zeta+\phi_0\,, 
\qquad
w^m = w_0^m\,,
\end{equation}
where $u_0$, $\phi_0$ and $w_0^m$ are constants. Clearly to close the $\phi$ circle (for $u_0\neq0$), we need to wrap the $\zeta$ circle $k$ times (or $k/2$ for even $k$).

$w_0^m$ correspond to the choice of scalars that are turned on in 
\eqref{Eq:Singularity}, and $u_0$ and $\phi_0$ are fixed by
\begin{equation}\label{Eq:ParameterMap}
\sinh u_0 = \frac{1}{\pi}\sqrt\frac{k}{2 N} |\beta_1|\,,
\qquad
\phi_0 = -\frac{4\pi \alpha}{k}\,.
\end{equation}
For simplicity, we set $w^m_0=\phi_0=0$. 

We note that in the limit $\beta_1\to 0$, we recover the M2-brane configuration 
dual to the 1/2 BPS Wilson loop \cite{Drukker:2009hy, Giombi:2023vzu}, 
though wrapped $k$ or $k/2$ times over the $\zeta$ circle, rather than once as 
for a fundamental Wilson loop.

The induced metric on the M2-brane is 
\begin{equation}\label{Eq:hClassical}
ds^2=\frac{R^2 \cosh^2u_0 }{4}h_0\,,
\qquad
h_0=h_{AdS_2} + \frac{4}{k^2} d\zeta^2 \,.
\end{equation}
Including the pullback of $C_3$ \eqref{Eq:C3}, 
the classical action $S^{(0)}$ is
\begin{equation}\label{Eq:ClassicalAction}
S^{(0)} = \frac{T_{M2} R^3}{4k} \int \Omega_{AdS_2} d\zeta 
= 
\begin{cases} 
- T_{M2} R^3 \pi^2\,, & k \text{ odd,}\\
- T_{M2} R^3 \pi^2 / 2\,, & k \text{ even,}
\end{cases}
\end{equation}
where $\Omega_{AdS_2}$ is the $AdS_2$ volume form, and we have used that the 
regularised volume for $AdS_2$ is $-2 \pi$. This gives the classical 
expectation values \eqref{Eq:ExpectationValue}. This is $k$ (for odd $k$) 
or $k/2$ (for even $k$) times the expectation value of the 1/2 BPS 
Wilson loop \cite{Drukker:2009hy, Marino:2009jd}.

\subsection{Bosonic fluctuations}

To compute the one-loop corrections, we need to integrate over quadratic fluctuations to the M2-brane action. At quadratic order, the bosonic and fermionic fluctuations decouple.

The bosonic part of the fluctuation action is obtained by expanding the action \eqref{classical action} around the classical solution \eqref{classical}. 
We take as world-volume coordinates $\sigma^a$ for $AdS_2$ and $\zeta$
and expand the other coordinates around their classical values
\begin{equation}\label{fluctuations}
u(\sigma^a,\zeta) = u_0 + \eta(\sigma^a,\zeta)\,, \qquad
\phi(\sigma^a,\zeta) =\frac{2}{k}\zeta + \varphi(\sigma^a,\zeta)\,, \qquad
w^m(\sigma^a,\zeta)\,.
\end{equation}
$\eta$, $\varphi$, $w^{m}$ and $\bar{w}^{\bar m}$ are the fluctuation fields.
The one-form $A$ in \eqref{Eq:OneForm} appearing in the background 
metric is to quadratic order
\begin{equation}
A \simeq \frac{i}{2}(w^m d\bar{w}^{\bar{m}} - \bar{w}^{\bar{m}} dw^m)\,.
\end{equation}
With this, the metric induced from \eqref{background metric} is to quadratic order
\begin{equation}\label{metric expansion}
\begin{aligned}
ds^2 \simeq 
\frac{R^2}{4}\Bigg(&\left(\cosh^2 u_0(1+2\eta^2) + 2\eta \sinh u_0\cosh u_0-\eta^2\right) h_0
\\
&+ \left(\partial_i \eta \partial_j \eta\, 
+ \sinh^2 u_0\, \partial_i \varphi\partial_j\varphi\, 
+ 4 \partial_iw^m \partial_j \bar{w}^{\bar{m}}\right) d\sigma^i d\sigma^j 
\\ &
+ \frac{4}{k}\left(\sinh^2 u_0\, \partial_i\varphi
+ (iw^m \partial_i \bar{w}^{\bar{m}}-i\bar{w}^{\bar{m}} \partial_i w^m)\right)d\sigma^i d\zeta 
\bigg)\,.
\end{aligned}
\end{equation}
This gives the quadratic correction to the Nambu-Goto action
\begin{equation}\label{Eq:Nambu-half-BPS}
\begin{aligned}
S_{\text{NG}}^{(2)}
&= \frac{R^3}{16}T_{\text{M2}}\cosh u_0\int \sqrt{h_0}
\bigg(h_{0}^{ij}(\partial_i \eta\partial_j \eta
+ \tanh^2u_0\,\partial_i\varphi\partial_j\varphi
+4\partial_i w^m \partial_j \bar{w}^{\bar{m}}) 
\\&\quad
+3(3\cosh^2 u_0-2)\eta^2 
+ k(3 \cosh^2u_0-1)\tanh u_0\, \eta \partial_\zeta \varphi 
+ \frac{ik}{4}  ( w^m \partial_\zeta \bar{w}^{\bar{m}} - \bar{w}^{\bar{m}} \partial_\zeta w^m)\bigg)\,.
\end{aligned}
\end{equation}
The pullback of the three form $C_3$ to quadratic order is
\begin{equation}\label{Eq:C3Expansion}
{}^* C_3^{(2)} 
= -\frac{R^3 \sqrt{h_0} \cosh u_0}{16} \left(
3 \cosh u_0\sinh u_0 \, \eta\partial_\zeta \varphi 
+ 3(3\cosh^2u_0-2)\eta^2\right).
\end{equation}

We define a further complex field $\chi$ as
\beq
\chi = \frac{1}{2}(\eta -i\tanh u_0\,\varphi)\,.
\eeq
Combining \eqref{Eq:Nambu-half-BPS} and \eqref{Eq:C3Expansion}, we obtain the quadratic 
bosonic action
\begin{equation}
\label{quad-action}
\begin{aligned}
S^{(2)}= \frac{R^3}{4}T_{\text{M2}}\cosh u_0\int d^3\sigma \sqrt{h_0}\bigg(&
h_{0}^{ij}(\partial_i\bar{\chi}\partial_j\chi
+\partial_i w^m \partial_j \bar{w}^{\bar{m}})
\\&
+\frac{ik}{4} (\chi \partial_\zeta \bar \chi - \bar \chi \partial_\zeta \chi
+w^m \partial_\zeta \bar{w}^{\bar{m}} - \bar{w}^{\bar{m}} \partial_\zeta w^m)\bigg)\,.
\end{aligned}
\end{equation}

$k$ dependence appears explicitly in the second line of \eqref{quad-action}, but 
also in the range of the $\zeta$ coordinate (see the comment under \eqref{classical}). 
Rescaling $\zeta$ to the range $[0,2\pi]$ replaces the $k$ in the second line 
by 1 or 2. For $u_0=0$ the resulting action is identical to that derived in 
\cite{Sakaguchi:2010dg} for M2-brane describing the 1/2 BPS Wilson loop 
in the $k=1$ and $k=2$ theories. The overall $\cosh u_0$ factor can 
be absorbed in a rescaling of the fluctuating fields and does not affect 
the determinant arising from the path integral. The result then agrees 
with the calculation of the determinant in \cite{Giombi:2023vzu}, which 
combined with the fermionic contribution gives the answer in \eqref{Eq:ExpectationValueOneLoop}.

\subsection{Fermionic fluctuations}
To write the action for the fermionic coordinates on the M2-brane, 
we need a vielbein for the target space.\footnote{We perform the calculation 
in Lorentzian signature, and Wick-rotate in the end.} 
A choice of vielbein (written as forms) is
\begin{equation}\label{vielbein}
R e^{\hat{a}} = \frac{\, R\cosh u}{2} \hat{e}^{\hat{a}} \,,
\quad 
R e^{\hat 2} =\frac{R \sinh u}{2}\, d\phi\,, 
\quad 
R e^{\hat 3} = \frac{R}{2}du\,,
\quad 
Re^{\hat 7}=\frac{R}{k}(d\zeta + k A)\,, 
\quad 
Re^{\hat{m}} \,,
\quad
Re^{\bar{\hat{m}}}\,,
\end{equation}
where $\hat{e}^{\hat{a}}$ are a vielbein of the unit radius $AdS_2$, and $e^{\hat{m}}$, $e^{\bar{\hat{m}}}$ are three pairs of complex 
conjugate vielbeine of ${\mathbb{CP}}^3$, written explicitly in \eqref{CP3 vielbeine}, such that
\begin{equation}\label{complex vielbein}
ds^2_{\mathbb{CP}^3} = \kappa_{\hat{m} \bar{\hat{n}}} e^{\hat{m}} e^{\bar{\hat{n}}} 
+ \kappa_{\bar{\hat{m}}{\hat{n}}} e^{\bar{\hat{m}}} e^{\hat{n}}\,,
\end{equation}
with $\kappa$ the K\"ahler metric on $\mathbb{C}^3$, explicitly 
$\kappa_{\hat{m} \bar{\hat{n}}} = \kappa_{\bar{\hat{m}} \hat{n}}= \delta_{mn}/2$, 
which is used to lower and raise the indices $\hat{m}$ and $\bar{\hat{m}}$.

The spin connection for this frame is \begin{equation}\label{EqSpinConnection}
\begin{aligned}
\Omega^{\hat 2}{}_{\hat 3}&= \cosh u\, d\phi\,, 
&\qquad 
\Omega^{\hat{a}}{}_{\hat 3}&= \sinh u\, \hat{e}^{\hat{a}}\,, 
&\qquad
\Omega^{\hat0}{}_{\hat1}\,,
&\qquad \\
\Omega^{\hat{m}}{}_{\hat{n}} &= 
- \Omega^{\bar{\hat{n}}}{}_{\bar{\hat{m}}} =
\hat{\Omega}^{\hat{m}}{}_{\hat{n}}
+i\delta^{\hat{m}}_{\hat{n}} e^{\hat7} \,,
&\qquad 
\Omega^{\hat7}{}_{\hat{m}}
&= K_{\hat{m} \bar{\hat{n}}}\, e^{\bar{\hat{n}}}\,,
&\qquad \Omega^{\hat7}{}_{\bar{\hat{m}}}
&= - K_{\hat{n} \bar{\hat{m}}} e^{\hat{n}}\,,
\end{aligned}
\end{equation}
where $\hat{\Omega}^{\hat{m}}{}_{\hat{n}}$ is the spin connection of 
unit radius $\mathbb{CP}^3$, \eqref{cp3 spin connection} \cite{Kihara:2008zg}. 
$\Omega^{\hat0}{}_{\hat 1}$ is the $AdS_2$ spin connection of the 
frame $\hat{e}^{\hat{a}}$, and
$K_{\hat{m}}{}_{\hat{n}}$ are the components of the K\"ahler two-form in this frame, i.e. 
$K=dA/2=K_{\hat{m} \bar{\hat{n}}} e^{\hat{m}}\wedge e^{\bar{\hat{n}}}$,
and are
\begin{equation}\label{Kcomponents}
K_{\hat{m} \bar{\hat{n}}} = \frac{i}{2}\delta_{\hat{m} \hat{n}}\,.
\end{equation}
The spin connection for $\mathbb{CP}^3$ with the $U(1)$ fibre is discussed in more detail in Appendix~\ref{CP3 appendix}.

The action for fermions at quadratic order involves a 32-component Majorana spinor $\theta$ 
and takes the form \cite{Bergshoeff:1987cm, deWit:1998yu, deWit:1998tk}
\begin{equation}\label{Fermionic action}
S_F^{(2)} = - \frac{T_{\text{M2}} R^2 \cosh^2 u_0}{4} 
\int d^3\sigma \sqrt{-h_0}\, \bar\theta \gamma^i (1-\gamma) D_i \theta\,,
\end{equation}
where $\bar \theta \equiv \theta^{\dagger}\Gamma_{\hat0}= \theta^\intercal C$, 
with $C$ being the charge conjugation matrix, $h_{0}$ is given in \eqref{Eq:hClassical}, 
and $\gamma^i$ are world-volume gamma matrices expressed in terms of the 11d matrices 
(after the rescaling in \eqref{Eq:hClassical}) as
\begin{equation}\label{EqWorldVolumeGammaMatricesDefinition}
\gamma^i = \frac{\cosh u_0}{2}\, {}^* e^{i}_{\hat{A}}\Gamma^{\hat{A}}\,,
\qquad
\gamma = \frac{1}{3!\sqrt{-h_{0}}} \epsilon^{ijk}\gamma_{ijk}\,.
\end{equation}
Here ${}^* e^{i}_{\hat{A}}$ is the pullback of the vielbein and 
$\epsilon^{ijk}$ is the Levi-Civita symbol with $\epsilon^{123}=1$.

Using the classical solution \eqref{classical} and our choice of vielbeine 
\eqref{vielbein}, the world-volume gamma matrices explicitly are
\begin{equation}\label{EqWorldVolumeMatricesHalf}
\gamma^{a}= \hat{e}^{a}_{\hat{b}}\, \Gamma^{\hat{b}}\,, 
\quad 
\gamma^\zeta = \frac{k}{2 \cosh u_0} \left(\Gamma^{\hat2} \sinh u_0 + \Gamma^{\hat7}\right),
\quad
\gamma = \frac{1}{\cosh u_0} \Gamma_{\hat0} \Gamma_{\hat1}\left(\Gamma_{\hat2} \sinh u_0 + \Gamma_{\hat7}\right),
\end{equation}
The 11d gamma matrices are those for the frame \eqref{vielbein} 
and satisfy the complexified Clifford algebra
\begin{equation}\label{gamma algebra}
\{\Gamma_{\hat{\mu}}, \Gamma_{\hat{\nu}}\} = 2 \eta_{\mu \nu}\,, 
\qquad 
\{{\Gamma}_{\hat{m}}, {\Gamma}_{\bar{\hat{n}}}\} = 2 \kappa_{\hat{m} \bar{\hat{n}}}\,, 
\qquad 
(\Gamma_{\hat7})^2=1\,,
\end{equation}
with all other anti-commutators vanishing. 
The world-volume ones then satisfy the usual anti-commutation relations 
\beq
\{ \gamma^i, \gamma^j\}=2h_{0}^{\,i j}\,.
\eeq

The covariant derivatives $D_i$ are defined as
\begin{equation}
\begin{aligned}\label{EqCovariantSpinors}
D_i &= \partial_i 
+ \frac{1}{4} {}^* {\Omega}_i^{\hat{A}\hat{B}}\Gamma_{\hat{A} \hat{B}}
- \frac{1}{2} \left(\,{}^*e_i^{\hat{\mu}} \Gamma^{\hat0\hat1\hat2\hat3} \Gamma_{\hat{\mu}} 
+ \, {}^*e_i^{\hat A} \Gamma^{\hat0\hat1\hat2\hat3} {\Gamma}_{\hat A}\right),
\end{aligned}
\end{equation}
noting that the second spin connection index is raised and we define $\Gamma_{\hat{m} \bar{\hat{n}}} = \frac{1}{2}\left(\Gamma_{\hat{m}}\Gamma_{\bar{\hat{n}}} - \Gamma_{\bar{\hat{n}}} \Gamma_{\hat{m}} \right)$ and similarly $\Gamma_{\bar{\hat{n}} \hat{m}}$. 
For our configuration, these are explicitly
\begin{equation}\label{Eq:covDerivatives} 
\begin{aligned}
D_{a} &= \partial_{a} - {}^*e_{a}^{\hat{a}} \Gamma^{\hat0\hat1\hat2\hat3} \Gamma_{\hat{a}} 
+ \frac{1}{2} {}^*\Omega_{a}^{\hat0\hat1}\Gamma_{\hat0\hat1} + \frac{\sinh u_0}{2} \,{}^*e^{\hat{a}}_{a}\Gamma_{\hat{a}\hat 3}\,, 
\\
D_\zeta &= \partial_\zeta -\frac{\sinh u_0}{2k}\Gamma^{\hat0\hat1\hat2\hat3}\, \Gamma_{\hat2} - \frac{1}{2k}\Gamma^{\hat0\hat1\hat2\hat3}\left(\Gamma_{\hat2} \sinh u_0 + \Gamma_{\hat7}\right) 
+ \frac{1}{2k}\left(2 \cosh u_0\Gamma_{\hat2\hat3}
+i \kappa^{\hat{m} \bar{\hat{n}}} \Gamma_{\hat{m}\bar{\hat{n}}}
\right).
\end{aligned}
\end{equation}

Substituting in the expression for the gamma matrices 
\eqref{EqWorldVolumeMatricesHalf} and the covariant derivatives 
\eqref{Eq:covDerivatives} leads to a somewhat complicated action. 
However, as was done in the context of strings in 
\cite{Drukker:2000ep, Forini:2015mca}, we can simplify it significantly 
by rotating the spinors such that the world-volume matrices $\gamma^i$ 
become constant. Such a rotation is equivalent to choosing an alternative 
vielbein to match the classical brane, as is done in Appendix~A.2 
of~\cite{Drukker:2023jxp} and also implemented for the 1/3 BPS vortex 
loop solution in Section~\ref{Sec:ThirdBPS} below. 

The world-volume matrix $\gamma^\zeta$ \eqref{EqWorldVolumeMatricesHalf} is proportional to $\Gamma_{\hat2} \sinh u_0+ \Gamma_{\hat7}$. We thus define a rotation matrix $S$
such that
\begin{equation}\label{transformation identities}
S^{-1}\Gamma_{\hat7} S = \frac{1}{\cosh u_0}\left(\Gamma_{\hat2} \sinh u_0 + \Gamma_{\hat7}\right), 
\qquad 
S^{-1} \Gamma_{\hat2} S = \frac{1}{\cosh u_0} \left(\Gamma_{\hat2} - \Gamma_{\hat7} \sinh u_0 \right).
\end{equation}
With this transformation, $S \gamma^\zeta S^{-1}$ is proportional to $ \Gamma_{\hat7}$
and $\gamma$ is transformed as $S\gamma S^{-1} = \Gamma_{\hat0\hat 1\hat 7}$. 
Defining the transformed spinor $\vartheta \equiv S\theta$, and imposing 
the kappa fixing equation
\begin{equation}\label{kappaFixing}
\frac{1}{2} \left(1 + \Gamma_{\hat0\hat1\hat7}\right)\vartheta = 0\,,
\end{equation}
we can replace $(1-\gamma)\vartheta=2\vartheta$ and 
find the fermionic action \eqref{Fermionic action} explicitly is
\begin{equation}\label{fermion action}
\begin{aligned}
S_F^{(2)} = -\frac{R^2 T_{\text{M2}}\cosh^2 u_0}{2}
\int d^3 \sigma \sqrt{-h_0} \,\bar{\vartheta} 
&\left( 
\left(\gamma^a \partial_{a} 
+ \frac{1}{2}  {}^*\Omega_{a}^{\hat0\hat1} \gamma^a\Gamma_{\hat0\hat1}\right) 
+ \frac{k}{2} \Gamma^{\hat7} \partial_\zeta \right. \\
&\quad \left. + \frac{5}{4} \Gamma^{\hat0\hat1\hat2\hat3} 
+ \frac{5}{2} \Gamma_{\hat3} \sinh u_0
+\frac{i}{4} \kappa^{\hat{m} \bar{\hat{n}}} \Gamma^{\hat7}\Gamma_{\hat{m}\bar{\hat{n}}}
\right)\vartheta\,,
\end{aligned}
\end{equation}
where the first line has the form of a kinetic term of a 3d spinor 
on $AdS_2\times S^1$ with metric $h_{0\, ab}$. 

The term proportional to $\Gamma_{\hat{3}}$ 
drops out from the action because of the kappa fixing condition. 
Thus, the dependence of the action on $u_0$ is only through the overall 
prefactor $\cosh^2 u_0$, similar to the case of the bosonic action \eqref{quad-action}, 
though the prefactor there is $\cosh u_0$.

To further simplify the mass term on the second line we use 
the relation for our representation of the gamma matrices  
$\Gamma^{\hat7} = 8 i \Gamma_{\hat0\hat1\hat2\hat3}
\Gamma_{\hat4 \bar{\hat4}} \Gamma_{\hat5\bar{\hat5}}\Gamma_{\hat6\bar{\hat6}}$
to write it as
\beq
-\frac{i}{4} \Gamma^{\hat7} \left(
40\Gamma_{\hat4\bar{\hat4}} \Gamma_{\hat5\bar{\hat5}} \Gamma_{\hat6\bar{\hat6}} 
+ 2\Gamma_{\hat4\bar{\hat4}} +2\Gamma_{\hat5\bar{\hat5}} +2\Gamma_{\hat6\bar{\hat6}}\right).
\eeq
We can easily diagonalise this matrix noting that 
$2 \Gamma_{\hat4 \bar{\hat4}}$, $2\Gamma_{\hat5 \bar{\hat5}}$ and $2\Gamma_{\hat6 \bar{\hat6}}$ 
all square to 1, commute with each other and are self-adjoint so that 
their eigenspinors are orthonormal and complete. For a simultaneous eigenspinor $\vartheta_{\alpha_4 \alpha_5 \alpha_6}$ of the three operators with 
eigenvalues $\alpha_4$, $\alpha_5$ and $\alpha_6$ respectively, this matrix 
evaluates to
\begin{equation}
-\frac{i}{4} \Gamma^{\hat7} 
\left(
40\Gamma_{\hat4\bar{\hat4}} \Gamma_{\hat5\bar{\hat5}} \Gamma_{\hat6\bar{\hat6}} 
+ 2\Gamma_{\hat4\bar{\hat4}} +2\Gamma_{\hat5\bar{\hat5}} +2\Gamma_{\hat6\bar{\hat6}}\right)\vartheta_{\alpha_4 \alpha_5 \alpha_6} 
=
- \frac{i}{4} \Gamma^{\hat7}
\left(5\alpha_4 \alpha_5 \alpha_6
+ \alpha_4 + \alpha_5 + \alpha_6\right) \vartheta_{\alpha_4 \alpha_5 \alpha_6}\,.
\end{equation}
The eigenvalues are then $\pm2i$ with degeneracy one and $\pm i$ with degeneracy 3.

This exactly matches the fermionic action found in \cite{Sakaguchi:2010dg} 
with $k=1$ for odd $k$ and $k=2$ for even $k$. The determinant of this differential 
operator was evaluated in~\cite{Giombi:2023vzu} and together with the bosonic part 
gives \eqref{Eq:ExpectationValueOneLoop}.

\section{The 1/3 BPS configuration}
\label{Sec:ThirdBPS}

The 1/2 BPS configuration \eqref{classical} describes an M2-brane embedded 
at a point in $\mathbb{CP}^3$. A generalisation of these solutions is obtained 
by extending the M2-brane along a circle in 
$\mathbb{CP}^1 \subset \mathbb{CP}^3$, 
leading to a 1/3 BPS configuration~\cite{Drukker:2008jm}. 
These represent vortex loops \eqref{Eq:Singularity} with non-zero $\beta_2$. 
They are extended along an $AdS_2\subset AdS_4$, whose coordinates we take 
as world-volume coordinates along with $\xi$, which is proportional 
to the phase of the $\mathbb{CP}^1$ coordinate $w^4$.%
\footnote{This is convenient in order to write the purely normal 
fluctuation coordinates in \eqref{thirdfluc}.}
The M2-brane has the following classical embedding coordinates
\begin{equation}\label{Eq:ThirdBPSWorldVolume}
\zeta = \xi+\zeta_0 \,,
\qquad
u = u_0\,,
\qquad 
\phi = \frac{2}{k}\xi  + \phi_0\,,
\qquad
w^4 = r e^{-2 i\xi / k}\,,
\qquad
\bar{w}^{\bar{4}} = r e^{2 i\xi/k}\,,
\end{equation}
$\xi$ has the range $[0, \pi k ]$ for even $k$, and $[0, 2 \pi k]$ if $k$ is odd. 
Closing in either case the $\zeta$, $\phi$ and $w_4$ circles.

The constants in \eqref{Eq:ThirdBPSWorldVolume} 
are related to the vortex loop parameters in \eqref{Eq:Singularity} by
\begin{equation}\label{Eq:Parameters}
\sinh u_0 =\frac{1}{\pi} \sqrt{\frac{k}{2 N}} \sqrt{|\beta_1|^2 + |\beta_2|^2}\,,
\qquad
\phi_0 = 2\pi \alpha\,,
\qquad
re^{-i\zeta_0}=\frac{\beta_2}{\beta_1}\,.
\end{equation}
As before, using the symmetries of the M2-brane action, we can set $\phi_0= \zeta_0 = 0$. 

With this solution, the pullback metric is the same as in the 1/2 BPS case \eqref{Eq:hClassical} and the classical action is the same as in \eqref{Eq:ClassicalAction} \cite{Drukker:2008jm}. We turn next to computing the action for the quadratic bosonic fluctuations around this classical solution.

\subsection{Bosonic fluctuations}
In order to have a simple action for the perturbations, 
we take the fluctuations to be normal to the classical world-volume 
and orthonormal to each-other. 
We denote the fluctuating fields 
$y_2$, $y_3$, $y_{m}$, and $\bar{y}_{\bar{m}}$ 
such that the embedding is
\begin{equation}
\label{thirdfluc}
\begin{aligned}
u &= u_0 + 2 y_3\,,
\\
\phi &= \frac{2}{k} \xi - \frac{2}{\sinh u_0\cosh u_0} y_2 \,, 
&\quad 
\zeta &= \xi + k \tanh u_0\, y_2 + \frac{ir k}{2} ( \bar{y}_{\bar 4} - y_4 ),
\\
w^4 &= e^{-2 i \xi/k} (r + y_4 + r^2 \bar{y}_{\bar 4} - 2 i r y_2),
&\quad 
\bar{w}^{\bar 4} &= e^{2 i \xi/k}(r + r^2 y_4 + \bar{y}_{\bar 4} + 2 i r y_2),
\\
w^{5} &= \sqrt{1 + r^2}\, y_5\,, 
&\quad 
\bar{w}^{\bar 5} &= \sqrt{1 + r^2}\,\bar{y}_{\bar 5}\,,
\\
w^{6} &= \sqrt{1 + r^2}\, y_6\,, 
&\quad 
\bar{w}^{\bar {6}} &= \sqrt{1 + r^2}\, \bar{y}_{\bar 6}\,,
\end{aligned} 
\end{equation}
For $r=0$ this matches the $1/2$ BPS case \eqref{fluctuations} with the 
identifications $y_4 e^{-2i\xi/k}\to w^4$, 
$\bar{y}_{4} e^{2 i \xi/k} \to\bar{w}^{\bar{4}}$ and other obvious maps. 

The pullback metric, written to second order in the perturbations is 
\begin{align}
ds^2 &\simeq \frac{R^2}{4} \Bigg( 
(\cosh^2 u_0(1+8y_3^2) + 4y_3\sinh u_0\cosh u_0 -4y_3^2) h_0 
\nonumber\\
&\quad{}+ 4 \left(\partial_i y_3 \partial_j y_3 + 4 \partial_i y_2 \partial_j y_2 + 4 \partial_i y_{m} \partial_j \bar{y}_{\bar{m}} \right) d\sigma^i d\sigma^j
+ \frac{4i}{k} \left( \bar{y}_{\bar{m}} \partial_i y_{m} - y_{m} \partial_i \bar{y}_{\bar{m}} + 8i y_3 \partial_i y_2\right)d\xi d\sigma^i 
\nonumber\\
&\quad{}+ \frac{4 r}{k} \left( ir \left( y_{4} \partial_i \bar{y}_{\bar{4}} - \bar{y}_{\bar 4} \partial_i y_4 \right)- 2\tanh u_0 \left( \left(y_4 + \bar{y}_{\bar 4} \right) \partial_i y_2 + \left(\partial_i y_4 + \partial_i \bar{y}_{\bar 4} \right) y_2 \right)\right) d\xi d\sigma^i\Bigg) \,,
\end{align}
where $h_0$ is defined in \eqref{Eq:hClassical}. We see that classically, the metric in the 1/2 BPS \eqref{metric expansion} and 1/3 BPS cases are the same.
The quadratic term in the Nambu-Goto action is
\begin{equation}
\begin{aligned}
S_{\text{NG}}^{(2)}
&= \frac{R^3}{4}T_{\text{M2}}\cosh u_0\int \sqrt{h_0}
\bigg(h_{0}^{ij}(\partial_i y_3 \partial_j y_3
+ \partial_i y_2 \partial_j y_2
+\partial_i y_{m} \partial_j \bar{y}_{\bar{m}}) \\
&\quad{}- 2 k y_3 \partial_\xi y_2 + 3(3\cosh^2u_0 - 2) y_3^2 
-2 k r \tanh u_0 \left( \left( y_4 + \bar{y}_{\bar 4} \right) \partial_\xi y_2 + \partial_\xi \left(y_4 + \bar{y}_{\bar 4} \right) y_2\right) \\
&\quad{}+ \frac{ik}{4} \left( y_{m} \partial_\xi \bar{y}_{\bar{m}} - \bar{y}_{\bar{m}} \partial_\xi y_{m} \right) + \frac{i k r^2}{4} \left(y_4 \partial_\xi y_4 + \bar{y}_{\bar 4}\partial_\xi \bar{y}_{\bar 4} \right)\bigg) \,.
\end{aligned}
\end{equation}
$r$ multiplies the last terms on both the second and third lines. Those are 
total $\xi$ derivatives and integrate to zero as the fluctuations are periodic. 
This means the dependence on $r$ drops out from the action.

The Wess-Zumino action is obtained by integrating the pulled back 3-form $C_3$ \eqref{Eq:C3}, obtaining at second order in the fluctuations
\begin{equation}
\label{1/3boson}
S_\text{WZ}^{(2)}
=\int {}^* C_3^{(2)} = \frac{R^3}{4}T_\text{M2}\cosh u_0\int d^3\sigma \sqrt{h_0} \left(
3 y_3^2 (3 \cosh^2 u_0 - 2) - 3 k y_3 \partial_\xi y_2 \right).
\end{equation}
Combining the two, we obtain the full quadratic bosonic action
\begin{equation}
\label{Eq:quad-action-third}
\begin{aligned}
S^{(2)}= \frac{R^3}{4}T_{\text{M2}}\cosh u_0\int d^3\sigma \sqrt{h_0}\bigg(
&h_{0}^{ij}\left(\partial_i y_{m}\partial_j\bar{y}_{\bar{m}}
+\partial_i y_3 \partial_j y_3 + \partial_i y_2 \partial_j y_2\right)
\\
&+k y_3 \partial_\xi y_2 + \frac{ik}{4} \left(y_{m} \partial_\xi \bar{y}_{\bar{m}} - \bar{y}_{\bar{m}} \partial_\xi y_{m} \right)\bigg)\,.
\end{aligned}
\end{equation}
Defining $\chi = y_3 + i y_2$, we obtain the same action for fluctuations 
in the 1/2 BPS case \eqref{quad-action}.

\subsection{Fermionic action}

To write the fermionic action in the 1/3 BPS case, we follow the logic 
as in Appendix~A.2 of \cite{Drukker:2023jxp} and choose a local frame 
at the classical solution \eqref{Eq:ThirdBPSWorldVolume} and adapted to it. 
We use $e^{\hat{a}}$, $e^{\hat3}$, $e^{\hat5}$, $e^{\bar{\hat5}}$, $e^{\hat6}$, 
$e^{\bar{\hat 6}}$ as in
\eqref{vielbein} and for the remaining four take
\begin{align}
e^{\hat7}&= \frac{\sinh u_0 \tanh u_0}{2} d\phi 
+ \frac{1}
{2k(1+ r^2)\cosh u_0}
\left(ik r (e^{2i\xi/k} dw^4 
-e^{-2i\xi/k}d\bar{w}^{\bar4})
-2(r^2-1)  d\zeta\right),
\nonumber\\
e^{\hat2}&= -\frac{\tanh u_0}{2} d\phi 
+ \frac{\tanh u_0}{2k (r^2+1)} 
\left(i kr (e^{2i \xi /k}dw^4 
- e^{-2i\xi/k} d\bar{w}^{\bar 4})
-2(r^2-1)  d\zeta\right), 
\label{Eq: thirdBPS-vielbein}
\\
e^{\hat4} &= \frac{e^{2 i \xi /k}}{1 + r^2} dw^4 
- \frac{2i r}{k (1 + r^2)} d\zeta\,,
\qquad
e^{\bar{\hat4}} =  \frac{e^{-2 i \xi /k}}{1 + r^2} d\bar{w}^{\bar 4} 
+ \frac{2i r}{k (1 + r^2)} d\zeta\,.
\nonumber
\end{align}
Note that $e^{\hat7}$ is tangential to the world-volume (as are $e^{\hat0}$ and $e^{\hat1})$ and 
$e^{\hat2}$, $e^{\hat4}$ and $e^{\bar {\hat4}}$ are normal. 

The fermionic action requires the pullback of the vielbeine and with this choice 
we only need to check the norm of $e_{\hat7}$, which pulls back to
\begin{equation}\label{Eq:thirdBPS-Pulledback-Vielbeine}
    {}^* e^{\xi}_{\hat7} = \frac{k}{\cosh u_0}\,.
\end{equation}
The world-volume gamma matrix $\gamma^\xi$, defined as in \eqref{EqWorldVolumeGammaMatricesDefinition}, then is
\begin{equation}
    \gamma^\xi =  \frac{k}{2} \Gamma^{\hat7}\,,
\end{equation}
while the other two $\gamma^a$ are as in the 1/2 BPS case \eqref{EqWorldVolumeMatricesHalf}. 

To complete the action we need to write the covariant derivatives 
\eqref{EqCovariantSpinors}, which requires the pullback of the 
spin connection. The definition in \eqref{Eq: thirdBPS-vielbein} is 
local and we cannot differentiate it to get the bulk spin connection. 
Instead, we get the pullbacks from the 11d Christoffel symbols 
for the metric~\eqref{background metric} $\Lambda^A{}_{BC}$ using
\begin{equation}
    {}^* \Omega_{i}^{\hat A \hat B} = t_i^C e_{A}^{\hat A}  \Lambda^{A}{}_{B C}\, e^{B \hat B} + e_{A}^{\hat A} \partial_{i} e^{B \hat B}\,,
\end{equation}
Here $t_i^C=\partial_i X^C$, with $X^C$ the target space coordinates 
\eqref{Eq:ThirdBPSWorldVolume}, are tangent vectors and in particular
\begin{equation}
t_\xi=\frac{2}{k} \partial_\phi + \frac{2ir}{k} \left( e^{-2i\xi/k} \partial_{w^4} - e^{2i \xi/k} \partial_{\bar{w}^{\bar 4}} \right) + \partial_\zeta\,.
\end{equation}

For the $AdS_2$ directions, ${}^*\Omega_{a}{}^{\hat A \hat B}$ are the same as 
found from \eqref{EqSpinConnection}. 
The components in the $\xi$ direction are
\begin{equation}\label{Eq:thirdBPS-Pulledback-Spinconnection}
\begin{aligned}
{}^*\Omega_{\xi}^{\hat{m} \bar{\hat n}} = i \kappa^{\hat m \bar{\hat n}}\,,
\qquad
{}^*\Omega_{\xi}^{\hat2\hat3} = \frac{2}{k}\,,
\qquad
{}^*\Omega_{\xi}^{\hat3\hat7} = \frac{2 \sinh u_0}{k}\,.
\end{aligned}
\end{equation}
Plugging this into the action \eqref{Fermionic action} and using the same 
kappa fixing condition as in \eqref{kappaFixing}
\begin{equation}\label{Eq:kappaFixing-ThirdBPS}
   \frac{1}{2} (1 + \Gamma_{\hat{1}\hat{2} \hat{7}})\theta =0\,,
\end{equation}
to replace $(1-\gamma)\theta=2\theta$, we immediately obtain 
\begin{equation}\label{Eq:Fermion-Action-ThirdBPS}
\begin{aligned}
S_F^{(2)} = -\frac{R^2 T_{\text{M2}}\cosh^2 u_0}{2}
\int d^3 \sigma \sqrt{h_0} \,\bar{\theta} 
&\left( 
\left(\gamma^a \partial_{a} 
+ \frac{1}{2} {}^*\Omega_{a}^{\hat0\hat1}\gamma^a \Gamma_{\hat0\hat1}\right) 
+ \frac{k}{2} \Gamma^{\hat7} \partial_\zeta \right. \\
&\quad \left. + \frac{5}{4} \Gamma^{\hat0\hat1\hat2\hat3} 
+  \frac{1}{4}\Gamma_{\hat3} \sinh u_0
+\frac{i}{4} \kappa^{\hat{m} \bar{\hat{n}}} \Gamma^{\hat7}\Gamma_{\hat{m}\bar{\hat{n}}}
\right)\theta\,,
\end{aligned}
\end{equation}
which is the same as 1/2 BPS case \eqref{fermion action}, apart from the 
coefficient of the $\Gamma_{\hat{3}}$ term, but it anyhow vanishes 
by the kappa fixing condition \eqref{Eq:kappaFixing-ThirdBPS}.

Both the bosonic \eqref{Eq:quad-action-third} and fermionic 
\eqref{Eq:Fermion-Action-ThirdBPS} actions agree with those of the 
1/2 BPS configuration \eqref{quad-action} \eqref{fermion action} 
and in turn with those of \cite{Sakaguchi:2010dg} in the $k=1$ and 
$k=2$ cases. The results of the one-loop determinant are then the 
same in all those examples, as discussed in Section~\ref{sec:intro}.

\section*{Acknowledgements}

We are grateful to S. Giombi, J. Maldacena and A. Tseytlin 
for helpful discussions. ND would like to thank 
CERN, EPFL and DESY 
for their hospitality in the course 
of this work. ND's research is supported by the Science 
Technology \& Facilities council under the grants 
and ST/P000258/1 and ST/X000753/1. 
OS's research is funded by the Engineering \& Physical Sciences Research Council under
grant number EP/W524025/1.

\appendix 

\section{Some properties of ${\mathbb{CP}}^3$ }\label{CP3 appendix}

In this appendix, we discuss some of the properties of the space ${\mathbb{CP}}^3$ including its vielbein basis and spin connection. The space $\mathbb{CP}^3$ can be parameterised by three complex coordinates $(w^4, w^5, w^6)$ with the Fubini-Study metric as written in \eqref{S7 metric}. For this metric, a choice of 3 complex vielbein, $e^{\hat{m}}$, 
with $\hat{m} = \hat4,\hat5,\hat6$, is \cite{Kihara:2008zg}
\begin{equation}\label{CP3 vielbeine}
e^{\hat{m}} = \frac{1}{\sqrt{1 + w^{\hat m} \bar{w}^{\bar{\hat m}}}} \left[dw^{\hat{m}} 
- \left(1 - \frac{1}{\sqrt{1 + w^{\hat m} \bar{w}^{\bar{\hat m}}}}\right) 
\frac{w^{\hat{m}}\bar{w}^{\bar{\hat{n}}} dw^{\hat{n}}}{w^{\hat m} \bar{w}^{\bar{\hat m}}} \right].
\end{equation}
The metric and K\"ahler form on unit radius $\mathbb{CP}^3$ 
is written in terms of the vielbeine as
\begin{equation}
ds_{\mathbb{CP}^3}^2 = \kappa_{\bar{\hat{m}} \hat{m}} e^{\bar{\hat{m}}}e^{\hat{m}} 
+
\kappa_{{\hat{m}} \bar{\hat{m}}} e^{{\hat{m}}}e^{\bar{\hat{m}}}\,,
\qquad
K = K_{\hat{m} \bar{\hat{n}}} e^{\hat{m}}\wedge e^{\bar{\hat{n}}}\,.
\end{equation}
where $\kappa_{\hat{m} \bar{\hat{n}}} 
= \kappa_{\bar{\hat{m}} \hat{n}}
=-iK_{\hat{m} \bar{\hat{n}}}= \delta_{mn}/2$.

The spin connection for $\mathbb{CP}^3$ in the basis $e^{\hat{m}}$ solves
\begin{equation}\label{spinConnectionComplex}
de^{\hat{m}} + \hat{\Omega}^{\hat{m}}{}_{\hat{n}} \wedge e^{\hat{n}} = 0\,,
\end{equation}
and is given by \cite{Kihara:2008zg}
\begin{equation}\label{cp3 spin connection}
\hat{\Omega}^{\hat{m}}{}_{\hat{n}} = i(\omega^{\hat{m}}{}_{\hat{n}} - A\, \delta^{\hat{m}}
_{\hat{n}})\,,
\end{equation}
where $\omega^{\hat{m}}{}_{\hat{n}}$ is real and is given by
\begin{equation}
\begin{aligned}
\omega^{\hat{m}}{}_{\hat{n}} &= \frac{1}{i |w|^2}
\left(1 - \frac{1}{\sqrt{1 + |w|^2}}\right) 
(w^{\hat{m}}d{\bar{w}}^{\bar{\hat{n}}} - \bar{w}^{\bar{\hat{n}}}dw^{\hat{m}}) 
\\&\quad
+ \frac{1}{2i |w|^4} 
\left(1 - \frac{1}{\sqrt{1 + |w|^2}}\right)^2
(\bar{w}^{\bar{\hat{k}}} dw^{\hat{k}} - w^{\hat{k}}d\bar{w}^{\bar{\hat{k}}})
w^{\hat{m}}\bar{w}^{\bar{\hat{n}}}\,,
\end{aligned}
\end{equation}
and $A$ is defined in \eqref{Eq:OneForm}.
We note that complex conjugation of \eqref{spinConnectionComplex} gives 
$\hat{\Omega}^{{\hat{m}}}{}_{{\hat{n}}}= - \hat{\Omega}^{\bar{\hat{n}}}{}_{\bar{\hat{m}}}$. 

\subsection{$U(1)$ Fibre over $\mathbb{CP}^3$}
The spin connection 
of the $U(1)$ fibre over $\mathbb{CP}^3$ 
solves
\begin{equation}
de^{\hat7} + \Omega^{\hat7}{}_{\hat{m}} \wedge e^{\hat{m}} 
+ \Omega^{\hat7}{}_{\bar{\hat{m}}} \wedge e^{\bar{\hat{m}}}=0\,.
\end{equation}
Choosing $e^{\hat7} = (d\zeta/k + A)$ as we did in \eqref{vielbein},
one has $de^{\hat7} = dA = 2 K = 2K_{\hat{m} \bar{\hat{m}}}e^{\hat{m}}\wedge e^{\bar{\hat{m}}}$ with the components $K_{\hat{m} \bar{\hat{n}}}$ as defined in \eqref{Kcomponents}. Therefore, we have
\begin{equation}
\Omega^{\hat7}{}_{\hat{m}}
= K_{\hat{m}\bar{\hat{n}}}\, e^{\bar{\hat{n}}}\,, \qquad \Omega^{\hat7}{}_{\bar{\hat{m}}}
= - K_{\hat{n} \bar{\hat{m}}}\, e^{\hat{n}}\,.
\end{equation}
Finally, $\mathbb{CP}^3$ components of the spin connection of the total space can be similarly found to be
\begin{equation}
\Omega^{\hat{m}}{}_{\hat{n}} = 
- \Omega^{\bar{\hat{n}}}{}_{\bar{\hat{m}}} =\hat{\Omega}^{\hat{m}}{}_{\hat{n}}
+i \delta^{\hat{m}}_{\hat{n}} e^{\hat7}\,.
\end{equation}

\bibliographystyle{utphys2}
\bibliography{mybibliography}

\end{document}